\documentstyle[11pt,aaspp4,psfig]{article}
\def\pp{\par\parshape 2 0truecm 15.5truecm 1truecm 14.5truecm\noindent}
\newcommand{\simgt}{\lower.5ex\hbox{$\; \buildrel > \over \sim \;$}}
\newcommand{\simlt}{\lower.5ex\hbox{$\; \buildrel < \over \sim \;$}}

\newcommand{\dm}{{\rm\scriptscriptstyle DM}}
\begin{document}

\title{Gas density and X-ray surface brightness profiles\\
of clusters of galaxies from dark matter halo potentials: \\
beyond the isothermal $\beta$ model
}

\bigskip

\author{Yasushi Suto$^{1,2}$, Shin Sasaki$^3$, and Nobuyoshi
Makino$^{4,5}$}

\bigskip
\bigskip

\affil{\altaffilmark{1} Department of Physics, University of Tokyo,
  Tokyo 113-0033, Japan} 
\affil{\altaffilmark{2} Research Center for the Early Universe
  (RESCEU), School of Science \\
  University of Tokyo, Tokyo 113-0033, Japan} 
\affil{\altaffilmark{3}  Department of Physics, 
  Tokyo Metropolitan University, Hachioji, Tokyo 192-0397, Japan}
\affil{\altaffilmark{4}  Department of Physics, 
  Ritsumeikan University, Kusatsu, Shiga, 525-8577, Japan}
\affil{\altaffilmark{5}  present address: Department of Mechanical 
Engineering \\ Oita National College of Technology,
1666 Maki, Oita 870-0152, Japan}

\bigskip

\affil{\footnotesize e-mail: suto@phys.s.u-tokyo.ac.jp,
sasaki@phys.metro-u.ac.jp,  makino@oita-ct.ac.jp}

\bigskip

\received{1998 March 9}
\accepted{1998 July 10}

\begin{abstract}
  We describe a theoretical framework to compute the cluster gas
  distribution in hydrostatic equilibrium embedded in a class of
  spherical dark matter halo potentials. Unlike the conventional
  isothermal $\beta$-model, the present method provides a physical
  basis to directly probe the shape of dark matter halo from the
  observed X-ray surface brightness and temperature profiles of
  clusters of galaxies.  Specifically, we examine the extent to which
  the resulting gas density and X-ray surface brightness profiles are
  sensitive to the inner slope of the dark matter halo density and
  other more realistic effects including the self-gravity of the gas
  and the polytropic equation of state. We also discuss a practical
  strategy to apply the present methodology to the actual cluster
  profiles from future X-ray observations.
\end{abstract}


\vspace{1cm}
\centerline{\it Accepted for publication in The Astrophysical Journal}

%%%%%%%%%%%%%%%%%%%%%%%%%%%%%%%%%%%%%%%%%%%%%%%%%%%%%%%%%%%%%%%%%%%%%%%
\clearpage

\section{Introduction}

The gas density profiles of X-ray clusters of galaxies are known to be
approximated well by the empirical formula, the isothermal
$\beta$-model:
%%%%%%%%%%%%%%%%%%%%%%%%%%%%%%%%%%%%%%%%%%%%%%%%%%%%%%%%%%%%%%%%%%
\begin{equation}
n_g(r) = {n_{g0} \over [1+(r/r_c)^2]^{3\beta/2}} .
\label{eq:betaprofile}
\end{equation}
%%%%%%%%%%%%%%%%%%%%%%%%%%%%%%%%%%%%%%%%%%%%%%%%%%%%%%%%%%%%%%%%%%
Theoretically this is consistent with the observed indication that
luminous member galaxies obey the King profile and the assumption of
the hydrostatic equilibrium of cluster gas. The galaxies in clusters,
however, constitute a very small fraction of the gravitational mass of
the entire cluster due to the presence of dark matter. Recent
high-resolution N-body/hydrodynamical simulations have strongly
suggested that dark halos of cluster scales are described by a family
of fairly universal density profiles. Navarro, Frenk \& White
(1996,1997, hereafter NFW) proposed a profile:
%%%%%%%%%%%%%%%%%%%%%%%%%%%%%%%%%%%%%%%%%%%%%%%%%%%%%%%%%%%%%%%%%%
\begin{equation}
\rho_\dm(r) = {\delta_c \rho_{c0} \over
(r/r_s)(1+r/r_s)^2 } ,
\label{eq:nfwprofile}
\end{equation}
%%%%%%%%%%%%%%%%%%%%%%%%%%%%%%%%%%%%%%%%%%%%%%%%%%%%%%%%%%%%%%%%%%
where $\rho_{c0}$ is the critical density of the universe at $z=0$,
and $\delta_c$ and $r_s$ are the concentration parameter and the
scaled radius whose explicit fitting formulae as a function of the
halo mass $M$ and cosmological parameters are found in Navarro, Frenk
\& White (1997).

This enables one to predict the profiles of the gas and the X-ray
surface brightness. Our previous work (Makino, Sasaki \& Suto 1998,
hereafter Paper I) examined the case of the NFW profile proposed, and
found that the resulting gas density profile is very close to the
empirical model (\ref{eq:betaprofile}).

Higher-resolution N-body simulations by Fukushige \& Makino (1997),
however, indicate that the inner density profile of the halo is much
steeper than the NFW profile above.  This conclusion was confirmed
later by a series of systematic N-body simulations by Moore et al.
(1998).  Evans \& Collett (1997) show that under some constraints the
density profile $\propto r^{-4/3}$ becomes the stationary solution to
the collisionless Boltzmann equation.  Furthermore, in Paper I we
assumed the isothermality and the possible effect of the temperature
profile was neglected.  The present paper explores these effects in
more details and provides several useful working formulae for the
X-ray surface brightness profiles as well.  In what follows we
describe various theoretical formulation and do not attempt to compare
with real observational data. The comparison with observations in this
context has been performed by Makino \& Asano (1998), Tamura (1998),
Xu et al. (1998) and Markevitch et al. (1998).  While all of their
results indicate that the current observational data are consistent
with the predictions in the universal density profile, the current
data quality in the spatial resolution of X-ray surface brightness and
temperature profiles is not sufficient in distinguishing from the
empirical $\beta$-model in an unambiguous manner. In this respect,
upcoming X-ray missions including AXAF and XMM should definitely
provide us the data suitable for that purpose.  This is mainly why we
do not attempt any tentative comparison with the currently available
data, but rather present various theoretical predictions which will be
useful for future data analysis.

The plan of the paper is as follows; in \S 2 we describe a new series
of analytical solutions for a family of density profiles of dark
matter halos generalizing Paper I. Then we examine effects of
self-gravity of the gas in \S 3, and non-isothermality by adopting
polytropic equation of state in \S 4. Finally \S 5 is devoted to
discussion and conclusions.

\section{Isothermal gas and X-ray surface brightness profiles
from a family of dark matter halo potentials}

We generalize the NFW profile (\ref{eq:nfwprofile}), and consider a
family of density profiles describing the dark matter halo:
%%%%%%%%%%%%%%%%%%%%%%%%%%%%%%%%%%%%%%%%%%%%%%%%%%%%%%%%%%%%%%%%%%
\begin{equation}
\rho_\dm(x) = {\delta_c \rho_{c0} \over x^\mu
(1+x^\nu)^\lambda },
\label{eq:haloprofile}
\end{equation}
%%%%%%%%%%%%%%%%%%%%%%%%%%%%%%%%%%%%%%%%%%%%%%%%%%%%%%%%%%%%%%%%%%
where $x \equiv r/r_s$ is the dimensionless radius in units of the
characteristic scale $r_s$.  Then the total mass of dark matter halo
within the radius $r$ is given by
%%%%%%%%%%%%%%%%%%%%%%%%%%%%%%%%%%%%%%%%%%%%%%%%%%%%%%%%%%%%%%%%%%
\begin{equation}
M(r) = 4\pi \delta_c \rho_{c0} r_s^3 m(r/r_s) ,
\label{eq:totmass}
\end{equation}
%%%%%%%%%%%%%%%%%%%%%%%%%%%%%%%%%%%%%%%%%%%%%%%%%%%%%%%%%%%%%%%%%%
with 
%%%%%%%%%%%%%%%%%%%%%%%%%%%%%%%%%%%%%%%%%%%%%%%%%%%%%%%%%%%%%%%%%%
\begin{equation}
m(x) \equiv \int_0^{x} {u^{2-\mu} \over (1+u^\nu)^\lambda} du .
\label{eq:mx}
\end{equation}
%%%%%%%%%%%%%%%%%%%%%%%%%%%%%%%%%%%%%%%%%%%%%%%%%%%%%%%%%%%%%%%%%%

If one neglects the gas and galaxy contributions to the gravitational
mass, the gas density profile $\rho_g(r)$ in hydrostatic equilibrium
with the above dark matter potential satisfies
%%%%%%%%%%%%%%%%%%%%%%%%%%%%%%%%%%%%%%%%%%%%%%%%%%%%%%%%%%%%%%%%%%%%%%%%%%%%%
\begin{equation}
  {k T_{\rm g} \over \mu_g m_p}{d \ln \rho_g \over dr} = - {G M(r) \over r^2} ,
\label{eq:equilib}
\end{equation}
%%%%%%%%%%%%%%%%%%%%%%%%%%%%%%%%%%%%%%%%%%%%%%%%%%%%%%%%%%%%%%%%%%%%%%%%%%%%%
where $\mu_g$ and $m_p$ denote the mean molecular weight of the gas
(we adopt 0.59 below) and the proton mass, and we assume that the gas
temperature $T_{\rm g}$ is constant over the cluster (non-isothermal
cases are considered in \S 4). Equation (\ref{eq:equilib}) can be
formally integrated to yield
%%%%%%%%%%%%%%%%%%%%%%%%%%%%%%%%%%%%%%%%%%%%%%%%%%%%%%%%%%%%%%%%%%%%%%%%%%%%%
\begin{equation}
\ln {\rho_g(r) \over \rho_{g0}} = -B \int_0^{r/r_s} {m(x) \over x^2} dx,
\label{eq:lnrhog}
\end{equation}
%%%%%%%%%%%%%%%%%%%%%%%%%%%%%%%%%%%%%%%%%%%%%%%%%%%%%%%%%%%%%%%%%%%%%%%%%%%%%
where
%%%%%%%%%%%%%%%%%%%%%%%%%%%%%%%%%%%%%%%%%%%%%%%%%%%%%%%%%%%%%%%%%%%%%%%%%%%%%
\begin{equation}
B \equiv {4\pi G\mu_g m_p\delta_c\rho_{c0}r_s^2 \over kT_{\rm g}} .
\label{eq:bdef}
\end{equation}
%%%%%%%%%%%%%%%%%%%%%%%%%%%%%%%%%%%%%%%%%%%%%%%%%%%%%%%%%%%%%%%%%%%%%%%%%%%%%
Note that $B$ is $27b/2$ in terms of the parameter $b$ defined in
Paper I. Also $B$ is rewritten as
%%%%%%%%%%%%%%%%%%%%%%%%%%%%%%%%%%%%%%%%%%%%%%%%%%%%%%%%%%%%%%%%%%%%%%%%%%%%%
\begin{equation}
 B = {3 \over \gamma m(1)} {T_{\rm vir}(r_s) \over T_{\rm g}} 
\label{eq:bastvir}
\end{equation}
%%%%%%%%%%%%%%%%%%%%%%%%%%%%%%%%%%%%%%%%%%%%%%%%%%%%%%%%%%%%%%%%%%%%%%%%%%%%%
in terms of the virial temperature defined as
%%%%%%%%%%%%%%%%%%%%%%%%%%%%%%%%%%%%%%%%%%%%%%%%%%%%%%%%%%%%%%%%%%%%%%%%%%%%%
\begin{equation}
T_{\rm vir}(r) \equiv
\gamma \frac{G \mu_g m_p M(r)}{3 k r} 
\label{eq:tvir}
\end{equation}
%%%%%%%%%%%%%%%%%%%%%%%%%%%%%%%%%%%%%%%%%%%%%%%%%%%%%%%%%%%%%%%%%%
with $\gamma (\approx 1 - 2) $ being a fudge factor (see Paper I).

With the density profile of the form (\ref{eq:haloprofile}), equation
(\ref{eq:lnrhog}) converges for $\mu<2$, and is rewritten as
%%%%%%%%%%%%%%%%%%%%%%%%%%%%%%%%%%%%%%%%%%%%%%%%%%%%%%%%%%%%%%%%%%%%%%%%%%%%%
\begin{eqnarray}
\label{eq:rhogprof}
\rho_g(r) &=& \rho_{g0} \exp[-Bf(r/r_s)] ,
\end{eqnarray}
%%%%%%%%%%%%%%%%%%%%%%%%%%%%%%%%%%%%%%%%%%%%%%%%%%%%%%%%%%%%%%%%%%%%%%%%%%%%
where
%%%%%%%%%%%%%%%%%%%%%%%%%%%%%%%%%%%%%%%%%%%%%%%%%%%%%%%%%%%%%%%%%%%%%%%%%%%%%
\begin{eqnarray}
\label{eq:fxdef}
f(x) &\equiv& \int_0^x {m(u) \over u^2} du 
= \int_0^x {u^{1-\mu} \over (1+u^\nu)^\lambda} du 
- {1 \over x}\int_0^x {u^{2-\mu} \over (1+u^\nu)^\lambda} du .
\end{eqnarray}
%%%%%%%%%%%%%%%%%%%%%%%%%%%%%%%%%%%%%%%%%%%%%%%%%%%%%%%%%%%%%%%%%%%%%%%%%%%%

In some specific cases, equations (\ref{eq:mx}) and
(\ref{eq:fxdef}) are analytically integrated:

\noindent (i) $\mu=1$, $\nu=1$, $\lambda=2$ (NFW)
%%%%%%%%%%%%%%%%%%%%%%%%%%%%%%%%%%%%%%%%%%%%%%%%%%%%%%%%%%%%%%%
\begin{eqnarray}
\label{eq:m1}
m(x) &=&  \ln(1+x) - {x \over 1+x} , \\
\label{eq:f1}
f(x) &=&  1 - {1 \over x} \ln(1+x).
\end{eqnarray}
%%%%%%%%%%%%%%%%%%%%%%%%%%%%%%%%%%%%%%%%%%%%%%%%%%%%%%%%%%%%%%%%

\noindent (ii) $\mu=3/2$, $\nu=3/2$, $\lambda=1$
%%%%%%%%%%%%%%%%%%%%%%%%%%%%%%%%%%%%%%%%%%%%%%%%%%%%%%%%%%%%%%%
\begin{eqnarray}
\label{eq:m2}
m(x) &=&  {2 \over 3}\ln(1+x^{2/3}) , \\
\label{eq:f2}
f(x) &=& {x-2\over 3x} \ln(1+\sqrt{x^3})
 + \ln(1+\sqrt{x})
 + {2 \over \sqrt{3}} \tan^{-1}{2\sqrt{x}-1 \over \sqrt{3}}
 + {\sqrt{3}\over 9}\pi .
\end{eqnarray}
%%%%%%%%%%%%%%%%%%%%%%%%%%%%%%%%%%%%%%%%%%%%%%%%%%%%%%%%%%%%%%%%

\noindent (iii) $\mu=3/2$, $\nu=1$, $\lambda=3/2$
%%%%%%%%%%%%%%%%%%%%%%%%%%%%%%%%%%%%%%%%%%%%%%%%%%%%%%%%%%%%%%%
\begin{eqnarray}
\label{eq:m3}
m(x) &=&  2 \ln(\sqrt{x}+\sqrt{1+x}) - 2 \sqrt{x\over 1+x} , \\
\label{eq:f3}
f(x) &=& 2 \sqrt{1+x\over x} 
- {2 \over x} \ln(\sqrt{x}+\sqrt{1+x}) .
\end{eqnarray}
%%%%%%%%%%%%%%%%%%%%%%%%%%%%%%%%%%%%%%%%%%%%%%%%%%%%%%%%%%%%%%%%

In what follows, we focus on the case with $\mu=\alpha$, $\nu=1$,
$\lambda=3-\alpha$ ($1<\alpha<2$). Then equation (\ref{eq:fxdef})
reduces to
%%%%%%%%%%%%%%%%%%%%%%%%%%%%%%%%%%%%%%%%%%%%%%%%%%%%%%%%%%%%%%%
\begin{equation}
f(x) = \int_0^x {u^{1-\alpha} \over (1+u)^{3-\alpha}} du 
- {1 \over x} \int_0^x 
{u^{2-\alpha} \over (1+u)^{3-\alpha}} du .
\label{eq:fx}
\end{equation}
%%%%%%%%%%%%%%%%%%%%%%%%%%%%%%%%%%%%%%%%%%%%%%%%%%%%%%%%%%%%%%%%
If we set $\alpha=1$, equation (\ref{eq:fx}) reduces to the case (i)
which corresponds to the original proposal by Navarro, Frenk \& White
(1996,1997) and is worked out in Paper I. Since the power-law slope of
the inner region is very sensitive to the mass resolution limit of
numerical simulations (Fukushige \& Makino 1997; Moore et al. 1998;
Melott et al. 1997; Splinter et al. 1998), we explore the similar
profiles by changing $\alpha$.  Since it is unlikely that the mass
resolution limit affects the asymptotic outer halo profile $\propto
r^{-3}$, we choose $\lambda=3-\alpha$ so as to reproduce the
asymptotic behavior.

\subsection{gas density profile}

\begin{figure}[tbp]
\begin{center}
%  \leavevmode\psfig{file=fxprof.ps,height=15cm}
  \leavevmode\psfig{file=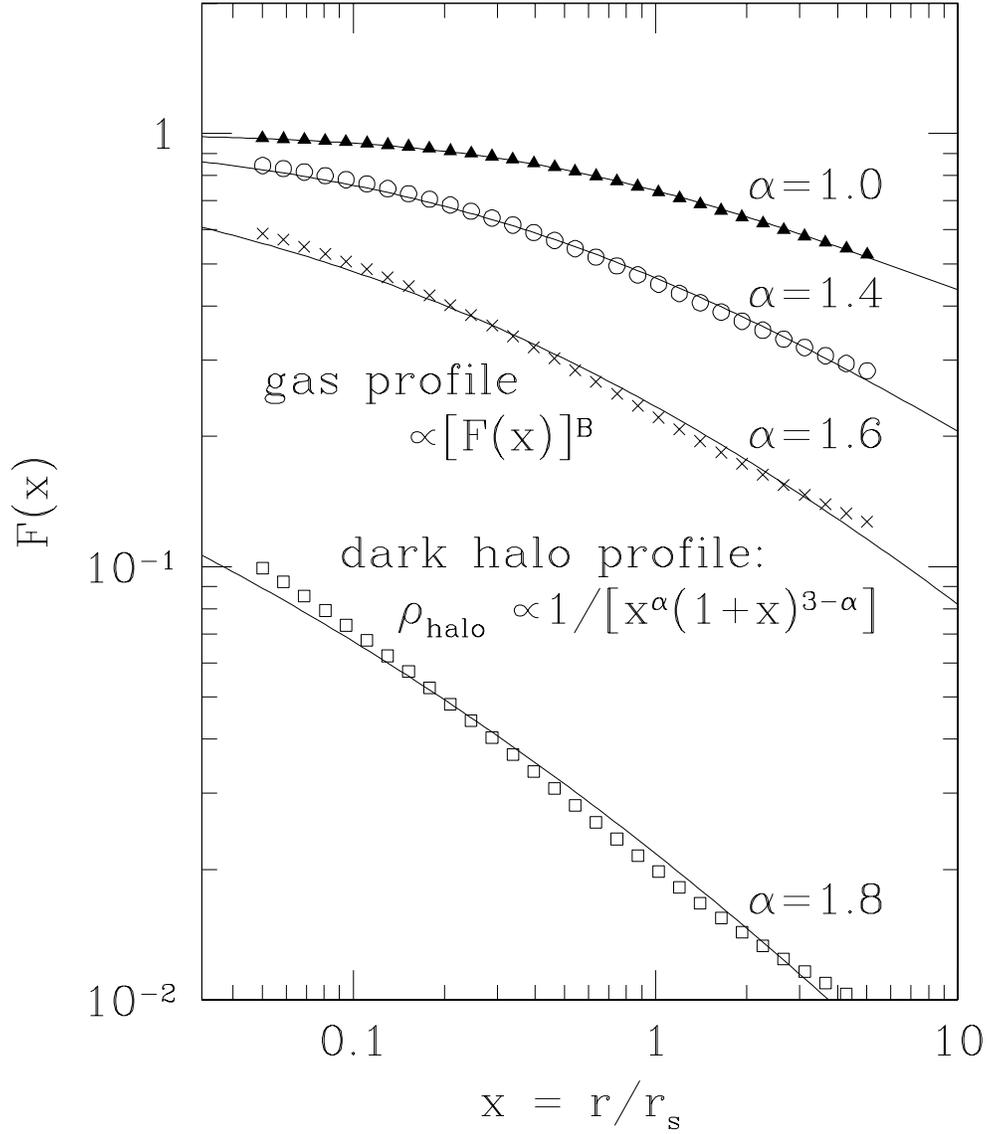,height=15cm}
\end{center}
\caption{Gas density profiles predicted from a family of 
dark matter halo profiles. Solid triangles, open circles, crosses, and
open squares indicate the results of numerical integrations for
$\alpha=1.0$, 1.4, 1.6, and 1.8, respectively. Solid lines represent
the best-fits to equation (\protect\ref{eq:fxfit}\protect) with 
using equation (\protect\ref{eq:xcpfit}\protect).
\label{fig:fxprof}}
\end{figure}

\begin{figure}[tbp]
\begin{center}
%  \leavevmode\psfig{file=fxfitparam.ps,height=15cm}
  \leavevmode\psfig{file=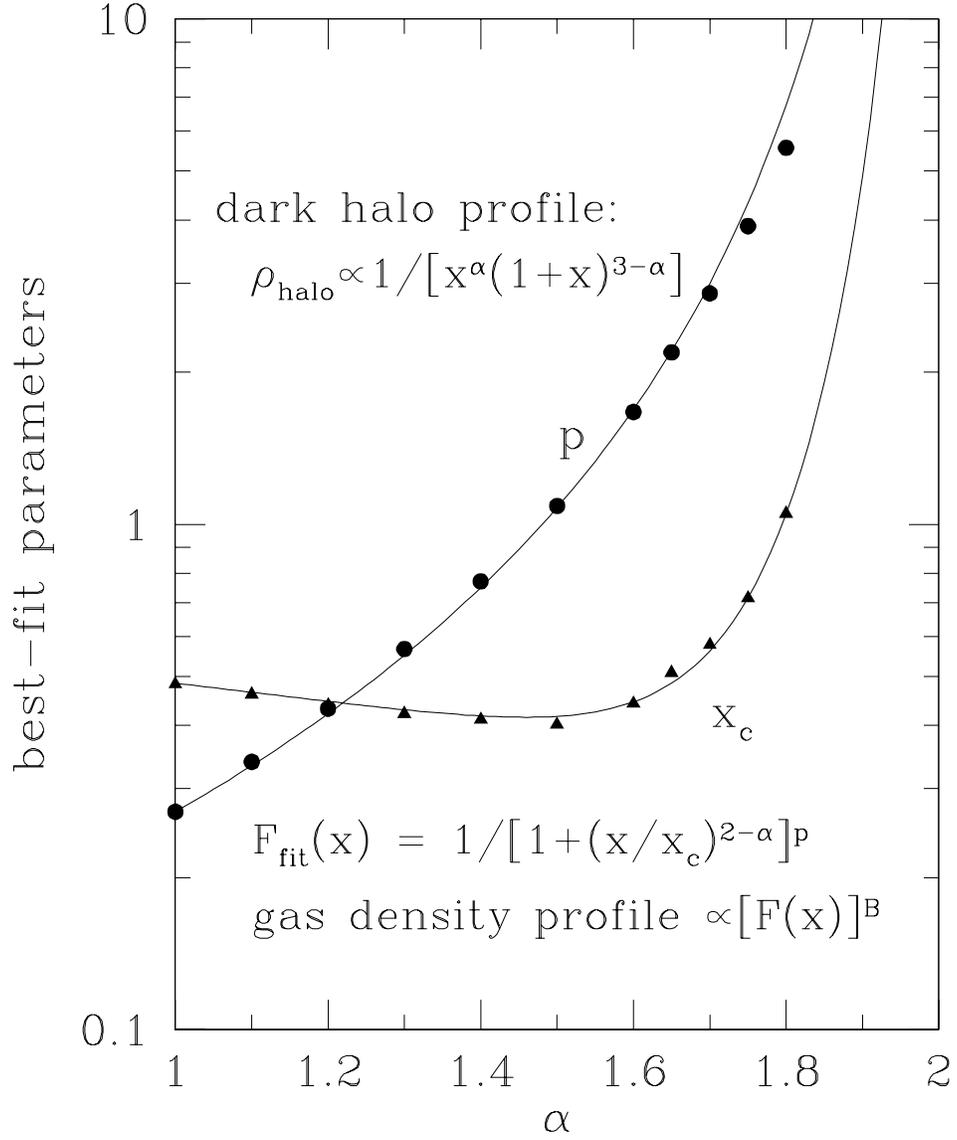,height=15cm}
\end{center}
\caption{The best-fit parameters of $x_c$ and $p$ as a function of
$\alpha$. Solid lines represent the fitting formula
(eq.[\protect\ref{eq:xcpfit}\protect]) which is
 accurate for $1.0 \leq \alpha \leq 1.6$.
\label{fig:fxfitparam}}
\end{figure}

For $1 \leq \alpha \leq 2$, we numerically integrate equation
(\ref{eq:fx}) to compute the gas density profile which is proportional
to $[F(x)]^B$ where we define
%%%%%%%%%%%%%%%%%%%%%%%%%%%%%%%%%%%%%%%%%%%%%%%%%%%%%%%%%%%%%%%%%%
\begin{equation}
F(x) \equiv \exp[-f(x)] .
\label{eq:largefx}
\end{equation}
%%%%%%%%%%%%%%%%%%%%%%%%%%%%%%%%%%%%%%%%%%%%%%%%%%%%%%%%%%%%%%%%%%
Figure \ref{fig:fxprof} plots the $F(x)$ for $\alpha=1.0$, $1.4$ and
$1.6$ together with the empirical fit to the following function:
%%%%%%%%%%%%%%%%%%%%%%%%%%%%%%%%%%%%%%%%%%%%%%%%%%%%%%%%%%%%%%%%%%
\begin{equation}
F_{\rm fit}(x) = \left[1+ \left({x \over
    x_c(\alpha)}\right)^{q(\alpha)} \right]^{p(\alpha)} .
\label{eq:fxfit}
\end{equation}
%%%%%%%%%%%%%%%%%%%%%%%%%%%%%%%%%%%%%%%%%%%%%%%%%%%%%%%%%%%%%%%%%%
For $x\ll1$, equations (\ref{eq:fx}) to (\ref{eq:fxfit}) are
consistent if
%%%%%%%%%%%%%%%%%%%%%%%%%%%%%%%%%%%%%%%%%%%%%%%%%%%%%%%%%%%%%%%%%%
\begin{equation}
q(\alpha) = 2 - \alpha
\label{eq:qalpha}
\end{equation}
%%%%%%%%%%%%%%%%%%%%%%%%%%%%%%%%%%%%%%%%%%%%%%%%%%%%%%%%%%%%%%%%%%
and 
%%%%%%%%%%%%%%%%%%%%%%%%%%%%%%%%%%%%%%%%%%%%%%%%%%%%%%%%%%%%%%%%%%
\begin{equation}
x_c^{2 - \alpha} = (3-\alpha)(2-\alpha)p .
\label{eq:xcalpha}
\end{equation}
%%%%%%%%%%%%%%%%%%%%%%%%%%%%%%%%%%%%%%%%%%%%%%%%%%%%%%%%%%%%%%%%%%
We adopt the relation (\ref{eq:qalpha}), but still keep $p$ and $x_c$
as two independent parameters to fit the $F(x)$ in the range of $0.05
\leq x \leq 5$. The results are plotted in solid lines in Figure
\ref{fig:fxprof}. For $1<\alpha<1.8$, we find the following
empirical fitting formulae:
%%%%%%%%%%%%%%%%%%%%%%%%%%%%%%%%%%%%%%%%%%%%%%%%%%%%%%%%%%%%%%%%%%
\begin{equation}
x_c(\alpha) = 0.015(2 - \alpha)^{-2.5}+ 0.47(2 - \alpha)^{0.5},
\qquad
p(\alpha) = 0.33(2-\alpha)^{-1.75} ,
\label{eq:xcpfit}
\end{equation}
%%%%%%%%%%%%%%%%%%%%%%%%%%%%%%%%%%%%%%%%%%%%%%%%%%%%%%%%%%%%%%%%%%
as plotted in Figure \ref{fig:fxfitparam}.

\subsection{X-ray surface brightness profile}

From the observational point of view, it is more useful to compute the
X-ray surface brightness profile on the sky:
%%%%%%%%%%%%%%%%%%%%%%%%%%%%%%%%%%%%%%%%%%%%%%%%%%%%%%%%%%%%%%%%%%
\begin{equation}
  \Sigma_{\rm X}(\theta) \equiv {1\over (1+z)^4}
  \int_{-\infty}^{\infty} \alpha_{\rm X}(T_{\rm g}) n_{\rm
  e}^2(\sqrt{\theta^2 d_A^2(z)+l^2})~dl ,
\label{eq:sx}
\end{equation}
%%%%%%%%%%%%%%%%%%%%%%%%%%%%%%%%%%%%%%%%%%%%%%%%%%%%%%%%%%%%%%%%%%
where $z$ is the redshift of the cluster considered, $\alpha_{\rm X}$
is the X-ray (either bolometric or band-limited) emissivity, $n_{\rm
e}(r)$ is the electron number density corresponding to the profile
(\ref{eq:rhogprof}), $d_A(z)$ is the angular diameter distance, and
the integration is performed over the line-of-sight at an angular
separation $\theta$ from the center of the cluster.

\begin{figure}[tbp]
\begin{center}
%  \leavevmode\psfig{file=sphiprof.ps,height=15cm}
  \leavevmode\psfig{file=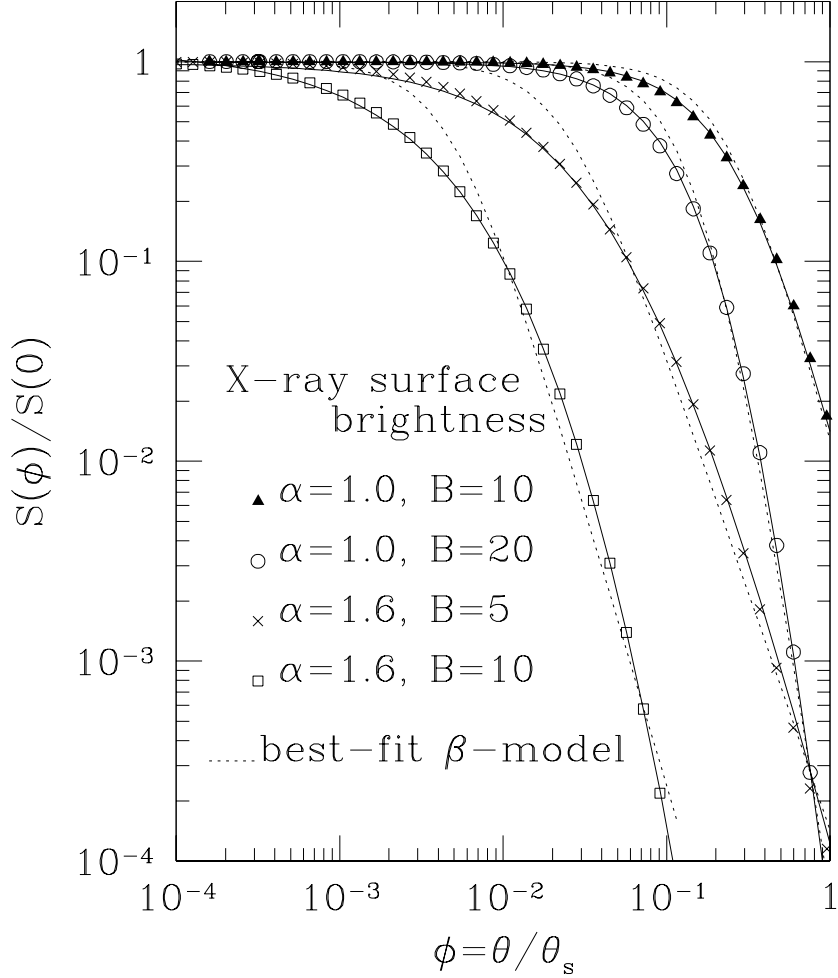,height=15cm}
\end{center}
\caption{X-ray surface brightness profiles
predicted from a family of dark matter halo profiles. Filled
triangles, open circles, crosses, and open squares indicate the
results of numerical integrations for $(\alpha,B)=(1.0,10.0)$, (1.0,
20.0), (1.6, 5.0), and (1.6, 10.0). Solid lines represent the
best-fits to equation (\protect\ref{eq:sfit}\protect) with equations
(\protect\ref{eq:paramfitphi}\protect) to
(\protect\ref{eq:paramfiteta}\protect), while dotted lines indicate
the best-fits to the conventional $\beta$-model
(eq.[\protect\ref{eq:sbetafit}\protect]).
\label{fig:sphiprof}}
\end{figure}

Substituting the isothermal density profiles (\ref{eq:fx}) and
(\ref{eq:largefx}), equation (\ref{eq:sx}) reduces to
%%%%%%%%%%%%%%%%%%%%%%%%%%%%%%%%%%%%%%%%%%%%%%%%%%%%%%%%%%%%%%%%%%
\begin{eqnarray}
\label{eq:sigmax}
\Sigma_{\rm X}(\theta) &=& {2\alpha_{\rm X}(T_{\rm g}) n_{\rm e0}^2 r_s
\over (1+z)^4}~ S(\theta/\theta_s) , 
\end{eqnarray}
%%%%%%%%%%%%%%%%%%%%%%%%%%%%%%%%%%%%%%%%%%%%%%%%%%%%%%%%%%%%%%%%%%
where $\theta_s \equiv r_s/d_A(z)$, and 
%%%%%%%%%%%%%%%%%%%%%%%%%%%%%%%%%%%%%%%%%%%%%%%%%%%%%%%%%%%%%%%%%%
\begin{eqnarray}
\label{eq:sphidef}
S(\phi) &\equiv& \int_\phi^\infty ~{x~\exp[-2Bf(x)] \over
\sqrt{x^2-\phi^2}} ~dx .
\end{eqnarray}
%%%%%%%%%%%%%%%%%%%%%%%%%%%%%%%%%%%%%%%%%%%%%%%%%%%%%%%%%%%%%%%%%%
Throughout the present paper we consider the X-ray bolometric
emissivity, and truncate the gas density at $r=20r_s$, i.e., $ n_g (r)
= 0 $ at $ r > 20 r_s$ in evaluating the surface brightness to avoid
unphysical divergence. This is because $20 r_s$ is typically larger
than the virial radius and the gas profile should not be extended at
the larger scale.  We integrate equation (\ref{eq:sphidef})
numerically for $(\alpha,B)=(1.0,5.0)$, (1.0, 10.0), (1.6, 10.0), and
(1.6, 20.0), and the results are plotted in Figure \ref{fig:sphiprof}.
They cannot be reasonably fitted to the conventional $\beta$-model:
%%%%%%%%%%%%%%%%%%%%%%%%%%%%%%%%%%%%%%%%%%%%%%%%%%%%%%%%%%%%%%%%%%
\begin{equation}
\label{eq:sbetafit}
S(\phi) = {S(0) \over [1+(\phi/\phi_{c,\beta})^2]^{3\beta-1/2}} ,
\end{equation}
%%%%%%%%%%%%%%%%%%%%%%%%%%%%%%%%%%%%%%%%%%%%%%%%%%%%%%%%%%%%%%%%%%%%
where $\beta$ is the slope parameter, $\phi_{c,\beta}$ is the core
radius, and $S(0)$ is the central surface brightness.  Due to the
strong concentration of the dark matter halo in the present models,
the resulting gas density profiles are also more concentrated than the
conventional $\beta$-model. In fact, the deviation from the
$\beta$-model becomes more appreciable for a larger $\alpha$.  Rather
we find that the following generalized shape:
%%%%%%%%%%%%%%%%%%%%%%%%%%%%%%%%%%%%%%%%%%%%%%%%%%%%%%%%%%%%%%%%%%
\begin{equation}
\label{eq:sfit}
S(\phi) \propto {1 \over [1+(\phi/\phi_c)^\xi]^\eta } 
\end{equation}
%%%%%%%%%%%%%%%%%%%%%%%%%%%%%%%%%%%%%%%%%%%%%%%%%%%%%%%%%%%%%%%%%%%%
with
%%%%%%%%%%%%%%%%%%%%%%%%%%%%%%%%%%%%%%%%%%%%%%%%%%%%%%%%%%%%%%%%%
\begin{eqnarray}
\label{eq:paramfitphi}
\phi_c&=&0.3\left(\frac{2}{\alpha}-1\right),\\
\label{eq:paramfitxi}
\xi&=&0.41-5.4(2-\alpha)^6+(0.585+6.47\alpha^{-5.18})B^{-\alpha^6/30},\\
\label{eq:paramfiteta}
\eta&=&-0.68-5.09(\alpha-1)^2+(0.202+0.0206\alpha^8)B^{1.1},
\end{eqnarray}
%%%%%%%%%%%%%%%%%%%%%%%%%%%%%%%%%%%%%%%%%%%%%%%%%%%%%%%%%%%%%%%%%
provides an excellent fit for $5\leq B\leq20$ and
$1.0\leq\alpha\leq1.6$ in the range of $10^{-4}\leq \phi \leq
\phi_{\rm max}$ where $S(\phi_{\rm max})=10^{-4}S(0)$.

Provided that the assumption of the isothermal gas is valid for actual
clusters, our fitting formulae above can be used directly to probe the
shape of the dark matter halo; fitting the observed X-ray surface
brightness to equations (\ref{eq:sfit}) to (\ref{eq:paramfiteta})
would result in the values of $\alpha$, $r_s$ and $B$ from the three
fitted parameters $\theta_s\phi_c$, $\xi$ and $\eta$. In addition if
the gas temperature $T_{\rm g}$ is measured from X-ray spectroscopic
observations, the concentration parameter $\delta_c$ is determined via
equation (\ref{eq:bdef}).

Incidentally let us note that our fitting formulae (\ref{eq:sfit}) are
not unique in the sense that the three parameters $\phi_c$, $\xi$
and $\eta$ are correlated and written in terms of the two independent
parameters $\alpha$ and $B$. In particular, $\phi_c$ and $\xi$ are
strongly correlated, and we would obtain an equally good fitting
formula by changing these two appropriately.  On the other hand,
$\eta$ is relatively insensitive to the choice of $\phi_c$ or $\xi$.

Applying the procedure described above to three lensing clusters
(A2163, A2218 and RX J1347.5-1145), Makino \& Asano (1998) showed that
the dark matter profile with $\alpha\leq 1.4$ is consistent with the
{\it ROSAT} HRI X-ray surface brightness profiles, and that $\alpha
\sim 1.4$ is preferred in order for the X-ray mass estimate to be
consistent with their giant arcs. Tamura (1998) also reached the
similar conclusion on the basis of ASCA/ROSAT observation of A1060.

\section{Effect of self-gravity of the gas density distribution}

When one properly includes the self-gravity of the gas distribution,
equation (\ref{eq:equilib}) reads
%%%%%%%%%%%%%%%%%%%%%%%%%%%%%%%%%%%%%%%%%%%%%%%%%%%%%%%%%%%%%%%%%%%%%%%%%
\begin{equation}
  {kT_{\rm g} \over \mu_g m_p}{d \ln \rho_g \over dr} 
= - {G M_{\rm tot}(r) \over r^2} 
= - {4\pi G \over r^2} \int_0^r u^2 [\rho_\dm(u)+\rho_g(u)]\,du .
\label{eq:gasgrav}
\end{equation}
%%%%%%%%%%%%%%%%%%%%%%%%%%%%%%%%%%%%%%%%%%%%%%%%%%%%%%%%%%%%%%%%%%%%%%%%%
With the profile (\ref{eq:haloprofile}) for dark matter halo, the
above equation is rewritten in a non-dimensional form as
%%%%%%%%%%%%%%%%%%%%%%%%%%%%%%%%%%%%%%%%%%%%%%%%%%%%%%%%%%%%%%%%%%
\begin{equation}
{d g(x) \over dx} = - {B \over x^2}\int_0^x 
\left[ {1\over u^\mu(1+u^\nu)^\lambda} + R e^{g(x)} \right]u^2 \, du ,
\label{eq:dgdx}
\end{equation}
%%%%%%%%%%%%%%%%%%%%%%%%%%%%%%%%%%%%%%%%%%%%%%%%%%%%%%%%%%%%%%%%%%
where we introduce 
%%%%%%%%%%%%%%%%%%%%%%%%%%%%%%%%%%%%%%%%%%%%%%%%%%%%%%%%%%%%%%%%%%
\begin{equation}
g(x) \equiv \ln[\rho_g(x)/\rho_{g0}]
\quad {\rm and} \quad
R \equiv  \rho_{g0}/(\delta_c \rho_{c0}).
\label{eq:grdef}
\end{equation}
%%%%%%%%%%%%%%%%%%%%%%%%%%%%%%%%%%%%%%%%%%%%%%%%%%%%%%%%%%%%%%%%%%
For $\mu=\alpha$, $\nu=1$ and $\lambda=3-\alpha$, the lowest-order
perturbation solution for equation (\ref{eq:dgdx}) at $x \ll 1$ is
%%%%%%%%%%%%%%%%%%%%%%%%%%%%%%%%%%%%%%%%%%%%%%%%%%%%%%%%%%%%%%%%%%
\begin{equation}
g(x) \approx  - {B \over (3-\alpha)(2-\alpha)}x^{2-\alpha} .
\label{eq:perturbg}
\end{equation}
%%%%%%%%%%%%%%%%%%%%%%%%%%%%%%%%%%%%%%%%%%%%%%%%%%%%%%%%%%%%%%%%%%

\begin{figure}[tbph]
\begin{center}
%  \leavevmode\psfig{file=rhomgasgravprof.ps,height=15cm}
  \leavevmode\psfig{file=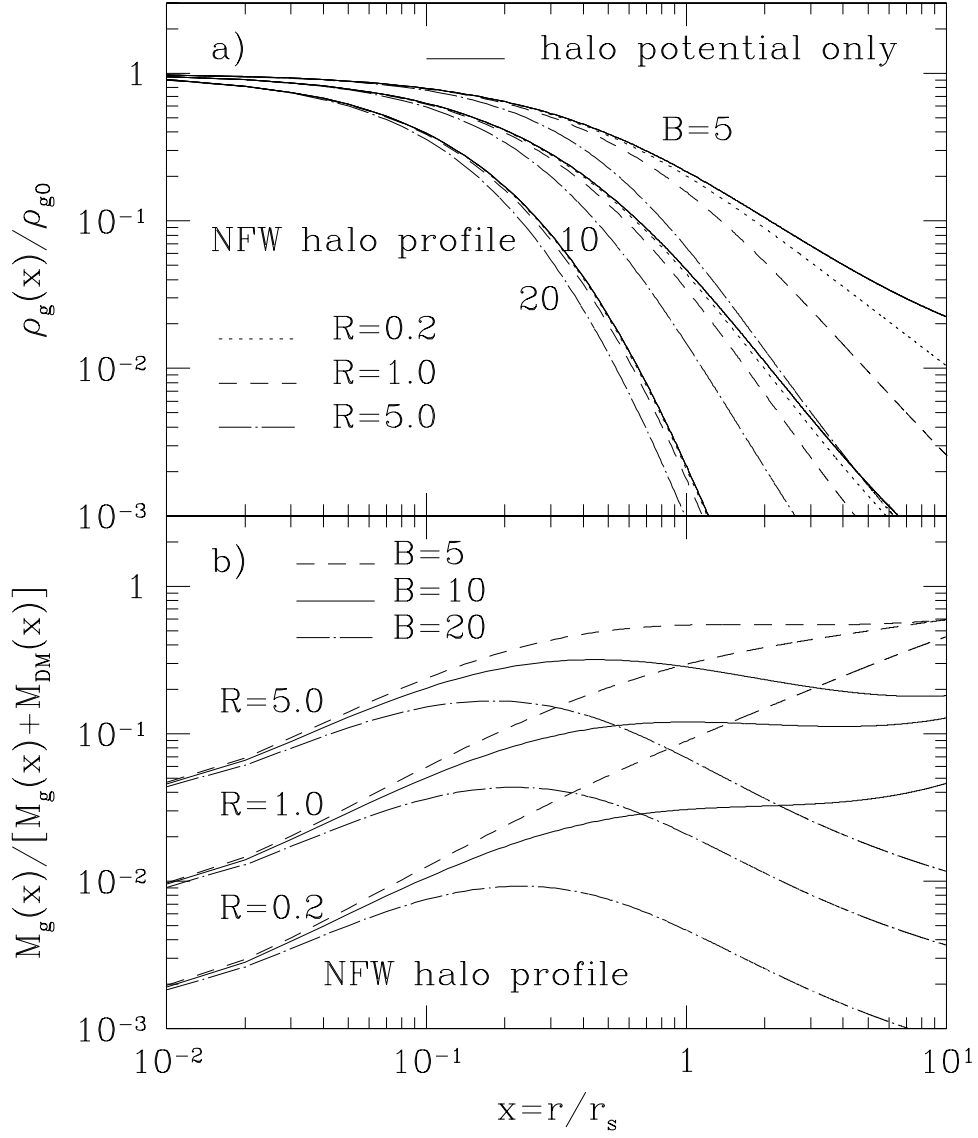,height=15cm}
\end{center}
\caption{Effect of self-gravity of gas density distribution;
a) gas density profile normalized in units of the central value;
b) gas mass fraction.  All models assume $\alpha=1$ (NFW profile).
\label{fig:rhomgasgravprof}}
\end{figure}

In Figure \ref{fig:rhomgasgravprof}, we consider the effect for the
NFW profile ($\alpha=1$) only, and numerically integrate the
second-order differential equation:
%%%%%%%%%%%%%%%%%%%%%%%%%%%%%%%%%%%%%%%%%%%%%%%%%%%%%%%%%%%%%%%%%%
\begin{equation}
{d^2 g(x) \over dx^2} +
{2 \over x}{d g(x) \over dx} 
+ B \left[ {1\over x(1+x)^2} + R e^{g(x)} \right] = 0 ,
\label{eq:d2gd2x}
\end{equation}
%%%%%%%%%%%%%%%%%%%%%%%%%%%%%%%%%%%%%%%%%%%%%%%%%%%%%%%%%%%%%%%%%%
where we rewrite equation (\ref{eq:dgdx}) with the boundary condition
specified from equation (\ref{eq:perturbg}).  The range of $R$ is
chosen so that the resulting gas to dark halo ratio:
%%%%%%%%%%%%%%%%%%%%%%%%%%%%%%%%%%%%%%%%%%%%%%%%%%%%%%%%%%%%%%%%%%
\begin{equation}
f_{g}(x) \equiv {M_g(x) \over M_\dm(x)} =
{R \over \ln(1+x) - x/(1+x)} \,\int_0^x e^{g(u)}u^2 \, du ,
\label{eq:fg10}
\end{equation}
%%%%%%%%%%%%%%%%%%%%%%%%%%%%%%%%%%%%%%%%%%%%%%%%%%%%%%%%%%%%%%%%%%
approximately ranges $10^{-3}\sim10^{-1}$ at $x=10$.

We plot the gas density profiles including the effect of self-gravity
in Figure \ref{fig:rhomgasgravprof}a. With increasing the gas fraction
or $R$, the gas density profile becomes steeper at large radii where
the self-gravity of gas becomes significant compared with that of dark
matter and confines the gas itself more strongly. Also as $B$
increases, the gas distribution becomes more centrally concentrated
since the gas temperature $T_g$ and therefore the pressure gradient
against the gravity becomes smaller for larger $B$.

In Figure \ref{fig:rhomgasgravprof}b we present the gas mass fraction
($= M_g/[M_g+M_\dm])$ for various values of $B$ and $R$.  At $x\simlt
1$, $\rho_{\dm}$ is proportional to $r^{-1}$ while $\rho_{\rm gas}$ is
approximately constant. So the gas fractions of clusters in the
present models increase roughly in proportional to $x$ in the inner
region. For larger $x$, on the other hand, the behavior is sensitive
to the values of $B$ and $R$.  The observed baryon fraction of
$0.1\sim 0.2$ of clusters indicate that $(B,R)=(10,5)$ and $(5,1)$
fall in an observationally relevant range.  In this case, the effect
of gas self-gravity is not significant, but cannot be fully neglected
either.  Also note that the gas mass fraction does not level off
anywhere.  Since the gas mass fraction provides an important
constraint on the density parameter $\Omega_0$ (e.g., White et
al. 1993), this indicates the need for the quantitative comparison on
the basis of the numerical integration of the above equation.  As
discussed in \S 6, this methodology is feasible under the
generalized halo potential model.

\section{Example of non-isothermal distribution: polytropic equation 
of state}

In the discussion above, we have assumed the isothermal gas
distribution.  Although this is regarded as a reasonable approximation
to the actual clusters, it is true that some clusters do show
non-isothermal features.  In particular, Markevitch et al. (1997)
reported that the temperature profiles of clusters appear remarkably
similar, and are approximately described by a polytrope with the
polytropic index $\Gamma = 1.2 - 1.3$ {\it assuming} that the gas
density profile is given by the $\beta$-model. In this section, we
consider the effect of non-isothermal gas distribution on the basis of
the polytropic equation of state for definiteness. More specifically,
we adopt the following form for the gas pressure $P$:
%%%%%%%%%%%%%%%%%%%%%%%%%%%%%%%%%%%%%%%%%%%%%%%%%%%%%%%%%%%%%%%%%%
\begin{equation}
\label{eq:ppolytrope}
  P = P_0(\rho_g/\rho_{g0})^{1+1/n} ,
\end{equation}
%%%%%%%%%%%%%%%%%%%%%%%%%%%%%%%%%%%%%%%%%%%%%%%%%%%%%%%%%%%%%%%%%%
or equivalently for the gas temperature:
%%%%%%%%%%%%%%%%%%%%%%%%%%%%%%%%%%%%%%%%%%%%%%%%%%%%%%%%%%%%%%%%%%
\begin{equation}
\label{eq:tpolytrope}
  T_{\rm g} = T_{\rm g 0}(\rho_g/\rho_{g0})^{1/n} ,
\end{equation}
%%%%%%%%%%%%%%%%%%%%%%%%%%%%%%%%%%%%%%%%%%%%%%%%%%%%%%%%%%%%%%%%%%
where the subscript 0 denotes the value at the center ($x=0$), and
the polytropic index $\Gamma$ is equal to $1 + 1/n$.

As before we assume that the gas is in hydrostatic equilibrium:
%%%%%%%%%%%%%%%%%%%%%%%%%%%%%%%%%%%%%%%%%%%%%%%%%%%%%%%%%%%%%%%%%%%%%%%%%%%%%
\begin{equation}
  {1 \over \rho_g}{d P \over dr} = - {G M(r) \over r^2} ,
\label{eq:pequilib}
\end{equation}
%%%%%%%%%%%%%%%%%%%%%%%%%%%%%%%%%%%%%%%%%%%%%%%%%%%%%%%%%%%%%%%%%%%%%%%%%%%%%
and neglect the self-gravity of the gas density. 
Define the function $\epsilon(x)$:
%%%%%%%%%%%%%%%%%%%%%%%%%%%%%%%%%%%%%%%%%%%%%%%%%%%%%%%%%%%%%%%%%%%%%%%%%%%%%
\begin{equation}
  \epsilon(x) \equiv [\rho_g(x)/\rho_{g0}]^{1/n} =
  T_{\rm g}(x)/T_{\rm g 0} ,
\label{eq:thetaxdef}
\end{equation}
%%%%%%%%%%%%%%%%%%%%%%%%%%%%%%%%%%%%%%%%%%%%%%%%%%%%%%%%%%%%%%%%%%%%%%%%%%%%%
then equation (\ref{eq:pequilib}) is written in a dimensionless form
as
%%%%%%%%%%%%%%%%%%%%%%%%%%%%%%%%%%%%%%%%%%%%%%%%%%%%%%%%%%%%%%%%%%
\begin{equation}
\label{eq:dthetadx}
  \frac{d\epsilon}{dx} = - B_{\rm p} {m(x) \over x^2} ,
\end{equation}
%%%%%%%%%%%%%%%%%%%%%%%%%%%%%%%%%%%%%%%%%%%%%%%%%%%%%%%%%%%%%%%%%%
where the constant $B_{\rm p}$ is defined as
%%%%%%%%%%%%%%%%%%%%%%%%%%%%%%%%%%%%%%%%%%%%%%%%%%%%%%%%%%%%%%%%%%
\begin{equation}
\label{eq:bp}
  B_{\rm p} \equiv \frac{4 \pi G}{n+1}
 \frac{\mu_g m_p \rho_{c0} \delta_c r_s^2 }{k T_{\rm g 0}} =
\frac{B}{n+1}.
\end{equation}
%%%%%%%%%%%%%%%%%%%%%%%%%%%%%%%%%%%%%%%%%%%%%%%%%%%%%%%%%%%%%%%%%%
Note that as in equation (\ref{eq:bastvir}) $B_{\rm p}$ is related to
$T_{\rm vir}(r_s)$ as
%%%%%%%%%%%%%%%%%%%%%%%%%%%%%%%%%%%%%%%%%%%%%%%%%%%%%%%%%%%%%%%%%%%%%%%%%%%%%
\begin{equation}
 B_{\rm p} = {3 \over (n+1)\gamma m(1)} 
  {T_{\rm vir}(r_s) \over T_{\rm g 0}} .
\label{eq:bpastvir}
\end{equation}
%%%%%%%%%%%%%%%%%%%%%%%%%%%%%%%%%%%%%%%%%%%%%%%%%%%%%%%%%%%%%%%%%%%%%%%%%%%%%
Equation (\ref{eq:dthetadx}) is integrated using the function $f(x)$
(eq.[\ref{eq:fxdef}]) to yield
%%%%%%%%%%%%%%%%%%%%%%%%%%%%%%%%%%%%%%%%%%%%%%%%%%%%%%%%%%%%%%%%%%%%%%%%%%%%%
\begin{equation}
  \epsilon(x) = 1 - B_{\rm p} f(x) .
\label{eq:thetax}
\end{equation}
%%%%%%%%%%%%%%%%%%%%%%%%%%%%%%%%%%%%%%%%%%%%%%%%%%%%%%%%%%%%%%%%%%%%%%%%%%%%%
Throughout the rest of this section we consider the NFW profile only
just for illustration, but the discussion below can be easily
generalized to other profiles.

\begin{figure}[tbph]
\begin{center}
%  \leavevmode\psfig{file=trhopolyprof.ps,height=15cm}
  \leavevmode\psfig{file=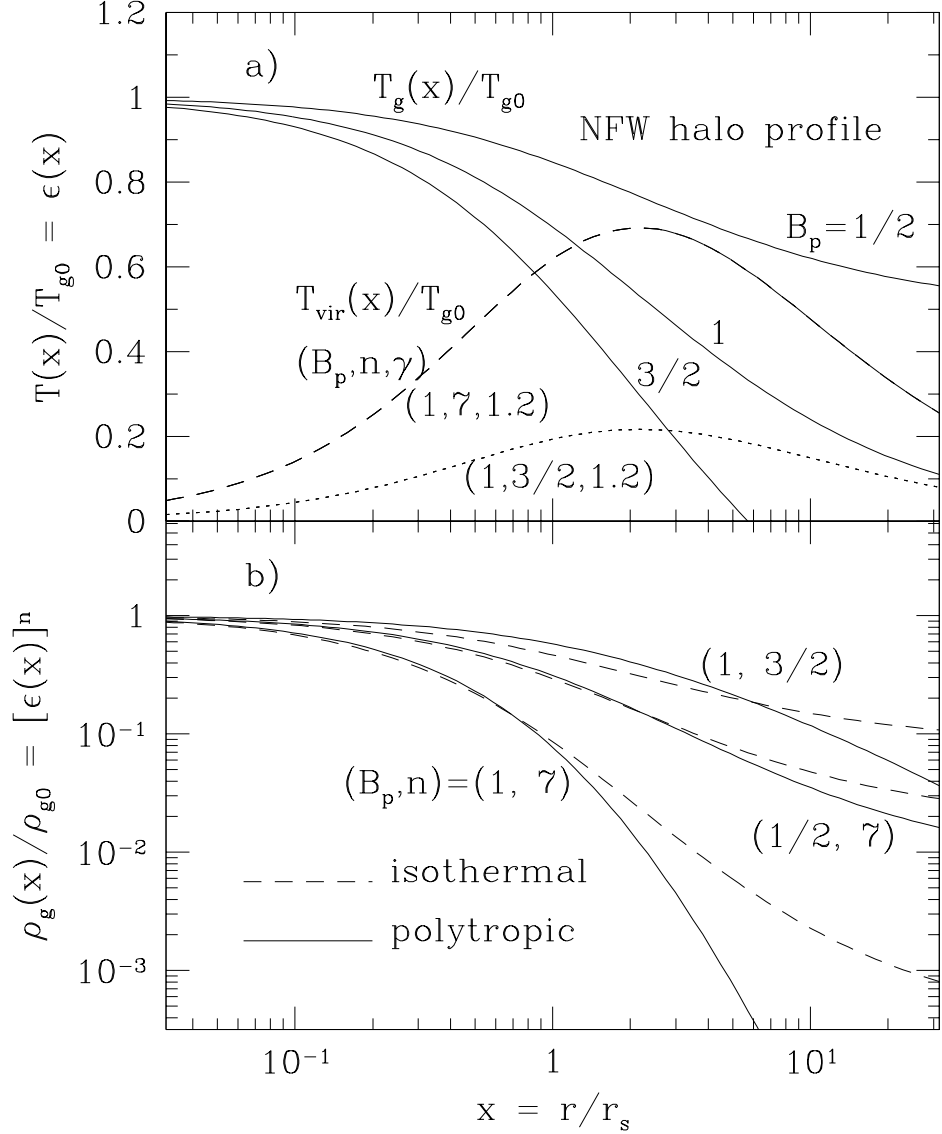,height=15cm}
\end{center}
\caption{ Profiles of temperatures and densities for polytropic models
embedded in the NFW dark matter halo potential ($\alpha=1$).  a) gas
temperatures $T_{\rm g}$ are plotted in solid lines for $B_p=0.5$,
$1.0$ and $1.5$. For reference, the virial temperature $T_{\rm vir}$
is plotted in dashed and dotted lines for $(B_p, n, \gamma) = (1.0, 7,
1.2)$ and (1.0, 3/2, 1.2), respectively. All values are in units of
the central gas temperature $T_{\rm g0}$; b) gas densities $\rho_{\rm
g}$ for polytropic models (solid lines) and for the corresponding
isothermal models (dashed lines) with $B=(n+1)B_p$ in units of the
central value $\rho_{\rm g0}$.
 \label{fig:trhopolyprof}}
\end{figure}

The temperature and gas profiles in the present polytropic model are
determined by specifying the two parameters, $B_{\rm p}$ and $n$.  It
should be noted that the upper limit on $B_{\rm p}$ is set by the
maximal extension of the cluster gas; if the gas extends up to
$x=x_{\rm max}$, then
%%%%%%%%%%%%%%%%%%%%%%%%%%%%%%%%%%%%%%%%%%%%%%%%%%%%%%%%%%%%%%%%%%%%%%%%%%%
\begin{equation}
\label{eq:bpmax}
  B_{\rm p} < B_{\rm p, max} \equiv 1/ f(x_{\rm max}) .
\end{equation}
%%%%%%%%%%%%%%%%%%%%%%%%%%%%%%%%%%%%%%%%%%%%%%%%%%%%%%%%%%%%%%%%%%%%%%%%%%%
In the case of the NFW profile, $B_{\rm p, max}$ are $1.32$, $1.18$
and $1$ for $x_{\rm max} = 10$, 20 and $\infty$, respectively.  With
equation (\ref{eq:bpastvir}), the above condition (\ref{eq:bpmax}) can
be translated to the lower limit on the central gas temperature as
%%%%%%%%%%%%%%%%%%%%%%%%%%%%%%%%%%%%%%%%%%%%%%%%%%%%%%%%%%%%%%%%%%%%%%%%%%%
\begin{equation}
\label{eq:tg0min}
T_{\rm g 0} > {3 T_{\rm vir}(r_s) \over (n+1)\gamma B_{\rm p, max}
m(1)} \sim {15 T_{\rm vir}(r_s) \over (n+1)\gamma} .
\end{equation}
%%%%%%%%%%%%%%%%%%%%%%%%%%%%%%%%%%%%%%%%%%%%%%%%%%%%%%%%%%%%%%%%%%%%%%%%%%%
The resulting temperature and gas profiles are plotted in Figure
\ref{fig:trhopolyprof}.  Note that $T_{\rm g}(x)/T_{\rm g 0}$ is
determined by $B_p$ and the halo shape parameters only and is
independent of the value of $n$, while the resulting gas density
profile is sensitive to $n$. The gas temperature in the present models
starts to decrease appreciably around at $x \simgt 1$ (solid lines in
Fig.\ref{fig:trhopolyprof}a), where the density profile departs
significantly from that in the corresponding isothermal model (dashed
lines in Fig.\ref{fig:trhopolyprof}b). The virial temperature of the
halo $T_{vir}(x)$ is significantly different from the gas temperature
$T_{\rm g}(x)$ in the present models (also in the isothermal model)
indicating the importance of the detailed numerical simulations (e.g.,
Eke, Cole \& Frenk 1998; Bryan \& Norman 1998; Yoshikawa, Itoh, \&
Suto 1998) in predicting the temperature profile of clusters even in
the present context.

\begin{figure}[tbph]
\begin{center}
%  \leavevmode\psfig{file=masspolyprof.ps,height=15cm}
  \leavevmode\psfig{file=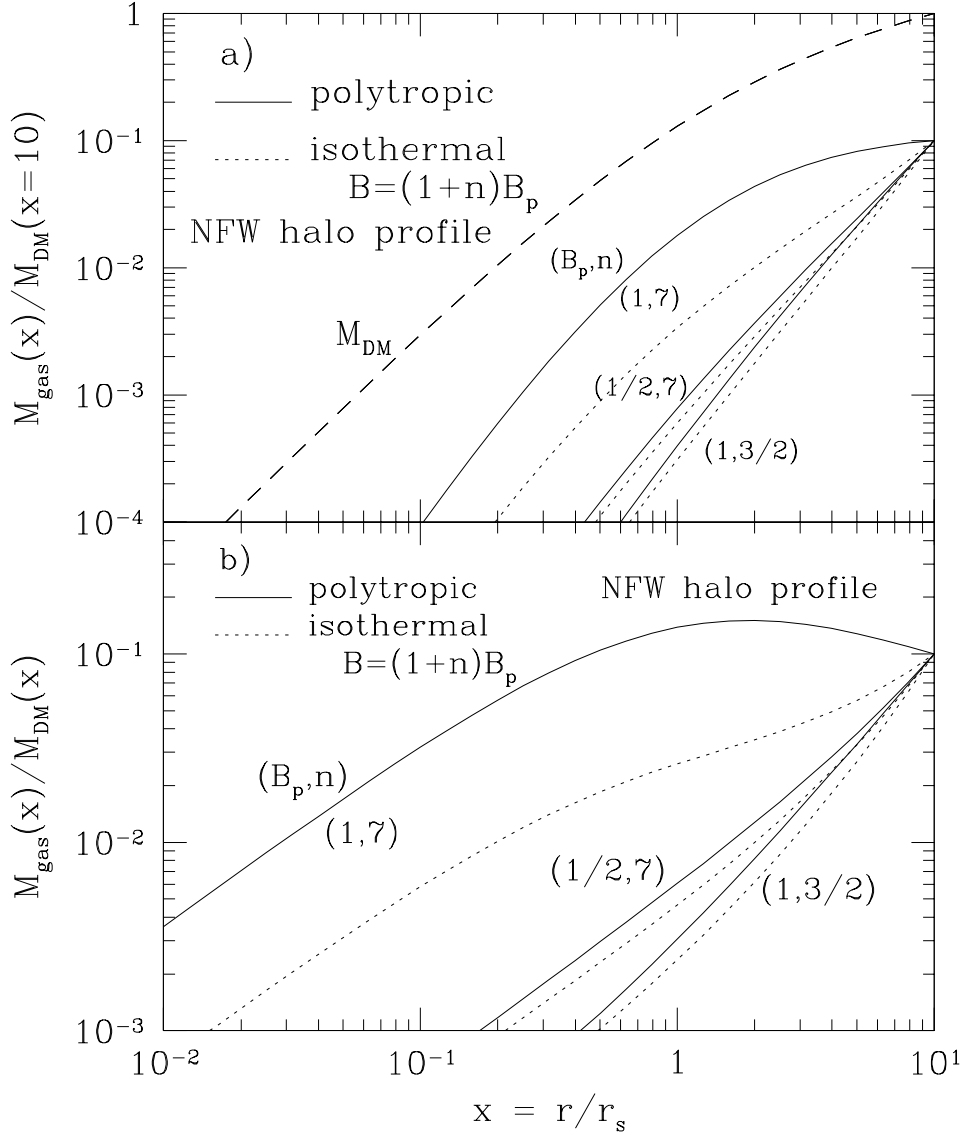,height=15cm}
\end{center}
\caption{ Mass profiles for polytropic models
embedded in the NFW dark matter halo potential ($\alpha=1$) with
$(B_p, n) = (1.0, 7.0)$, $(0.5,7.0)$, and (1.0,1.5).  a) gas mass
profiles for polytropic models (solid lines) and for the corresponding
isothermal models (dotted lines) with $B=(n+1)B_p$.  The gas mass is
normalized to be 10 percent of the mass of dark matter halo at $x=10$.
The halo mass profile is plotted in dashed line for reference; b) gas
to dark halo mass ratio for polytropic models (solid lines) and for
the corresponding isothermal models (dotted lines) with $B=(n+1)B_p$.
 \label{fig:masspolyprof}}
\end{figure}

Figure \ref{fig:masspolyprof} shows the mass profile and gas mass
fraction for polytropic models in comparison with those for isothermal
cases.  The gravity of gas mass is neglected, and for definiteness we
choose the gas mass to be 10 percent of that of the dark halo at
$x=10$ which roughly corresponds to the virial radius in the NFW
model.

\begin{figure}[tbph]
\begin{center}
%  \leavevmode\psfig{file=sphitpolyprof.ps,height=15cm}
  \leavevmode\psfig{file=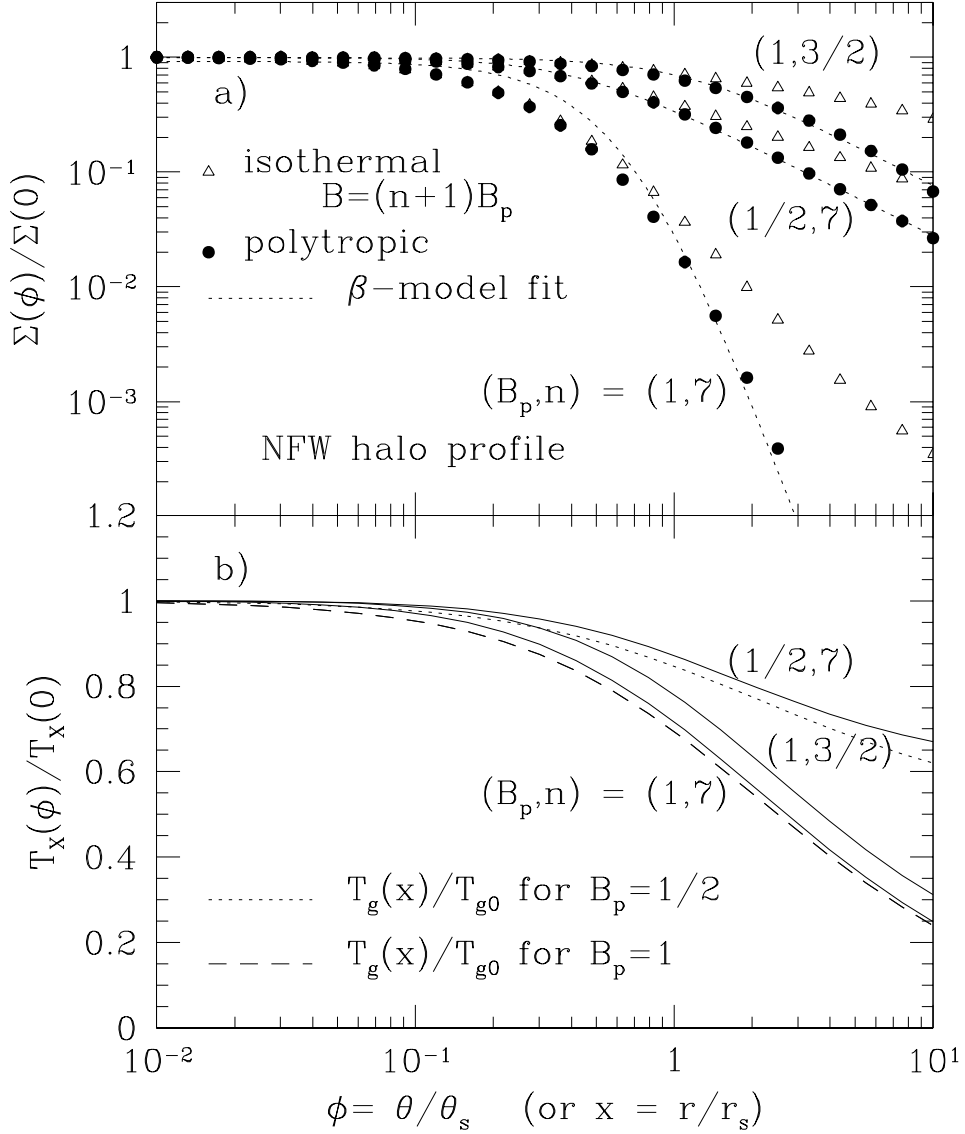,height=15cm}
\end{center}
\caption{X-ray surface brightness and 
and emission-weighted projected temperature profiles for polytropic
models embedded in the NFW dark matter halo potential ($\alpha=1$)
with $(B_p, n) = (1.0, 7.0)$, $(0.5,7.0)$, and (1.0,1.5).  a)
bolometric X-ray surface brightness (thermal bremsstrahlung only)
profiles for polytropic models (filled circles) and for the
corresponding isothermal models (open triangles) with $B=(n+1)B_p$.
The dotted lines represent the best-fits to the $\beta$ model.  b)
emission-weighted projected temperature profiles for polytropic models
(solid lines), gas temperature with $B_p=1/2$ (dotted line) and
$B_p=1$ (dashed line),
 \label{fig:sphitpolyprof}}
\end{figure}

The X-ray surface brightness and the emission-weighted temperature
profiles on the sky, $\Sigma_{\rm X}(\theta)$ and $T_{\rm X}(\theta)$,
are defined by equation (\ref{eq:sx}) and
%%%%%%%%%%%%%%%%%%%%%%%%%%%%%%%%%%%%%%%%%%%%%%%%%%%%%%%%%%%%%%%%%%
\begin{equation}
  T_{\rm X}(\theta) \equiv 
{
{\displaystyle 
\int_{-\infty}^{\infty} T_{\rm g}\alpha_{\rm X}(T_{\rm g}) n_{\rm
  e}^2(\sqrt{\theta^2 d_A^2(z)+l^2})~dl }
   \over 
{\displaystyle \int_{-\infty}^{\infty} \alpha_{\rm X}(T_{\rm g}) n_{\rm
  e}^2(\sqrt{\theta^2 d_A^2(z)+l^2})~dl }
} ,
\label{eq:tx}
\end{equation}
%%%%%%%%%%%%%%%%%%%%%%%%%%%%%%%%%%%%%%%%%%%%%%%%%%%%%%%%%%%%%%%%%%
respectively. If we consider the bolometric thermal bremsstrahlung
emissivity only, the polytropic models which we described above yield
%%%%%%%%%%%%%%%%%%%%%%%%%%%%%%%%%%%%%%%%%%%%%%%%%%%%%%%%%%%%%%%%%%
\begin{eqnarray}
\label{eq:polysx}
{\Sigma_{\rm X}(\phi) \over \Sigma_{\rm X}(0)}
&=& {\displaystyle \int_\phi^\infty ~{x~[\epsilon(x)]^{2n+1/2} \over
\sqrt{x^2-\phi^2}} ~dx
\over \displaystyle \int_0^\infty ~[\epsilon(x)]^{2n+1/2} ~dx } ,\\
\label{eq:polytx}
{T_{\rm X}(\phi) \over T_{\rm X}(0)}
&=& 
{\displaystyle \int_\phi^\infty ~{x~[\epsilon(x)]^{2n+3/2} \over
\sqrt{x^2-\phi^2}} ~dx 
\over 
\displaystyle \int_\phi^\infty ~{x~[\epsilon(x)]^{2n+1/2} \over
\sqrt{x^2-\phi^2}} ~dx } ~
{\displaystyle \int_0^\infty ~[\epsilon(x)]^{2n+1/2} ~dx 
\over \displaystyle \int_0^\infty ~[\epsilon(x)]^{2n+3/2} ~dx } ,
\end{eqnarray}
%%%%%%%%%%%%%%%%%%%%%%%%%%%%%%%%%%%%%%%%%%%%%%%%%%%%%%%%%%%%%%%%%%
where $\phi=\theta/\theta_s$ (\S 2.2). These profiles are plotted in
Figure \ref{fig:sphitpolyprof}. The dotted curves in the upper panel
indicate the best-fit $\beta$-model to the surface brightness. It is
somewhat surprising that the X-ray surface brightness profiles for
some models, e.g., with $(B_p,n)=(1.0,7.0)$, are well approximated by
the $\beta$-model despite the fact that the cluster is far from
isothermal. It should be also noted here that the projected
emission-weighted temperature $T_{\rm X}(\theta)$ systematically
differs from the gas temperature $T_{\rm g}(r)$ evaluated at
$\theta=r/d_{\rm A}$ (e.g., Figs.\ref{fig:trhopolyprof}a and
\ref{fig:sphitpolyprof}b).

\section{Discussion and conclusions}

In this paper, we have presented a physical methodology to confront
the dark matter halo mass distribution with the observed X-ray surface
brightness profiles of the galaxy cluster. Unlike the previous
phenomenological prescriptions including the isothermal $\beta$-model,
this approach enables one to determine the dark matter halo profile
directly from the observational data. This works in an especially
straightforward manner when the cluster is well approximated as
isothermal; then our fitting formulae (eqs.[\ref{eq:xcpfit}],
[\ref{eq:paramfitxi}] to [\ref{eq:paramfiteta}]) explicitly links the
gas density and X-ray surface brightness profiles to the underlying
dark matter halo potential as long as the halo is described by a
family of profiles (eq.[\ref{eq:haloprofile}]).  In \S 4, we have
described a prescription of computing the gas, temperature, and X-ray
surface brightness profiles for clusters with polytropic equation of
state. We confirm that the self-gravity of the gas does not affect the
density profile significantly (\S 3) in the halo profiles considered
throughout the paper. Incidentally Bardelli et al. (1996) claimed that
the gas mass fraction in A3558 (Shapley 8) is $\sim0.7$ at the Abell
radius (assuming $H_0=50{\rm km\,s^{-1} Mpc^{-1}}$). If confirmed, the
X-ray surface brightness profile for such clusters should be computed
with properly taking account of the self-gravity of gas, which is in
fact feasible as shown below.

When both the self-gravity of gas density and the polytropic equation
of state are taken into account, our procedure described in the
present paper is generalized as follows;

(i) select a dark matter halo density profile parameterized by
$(\mu, \nu, \lambda)$:
%%%%%%%%%%%%%%%%%%%%%%%%%%%%%%%%%%%%%%%%%%%%%%%%%%%%%%%%%%%%%%%%%%
\begin{equation}
\rho_\dm(r) = {\delta_c \rho_{c0} \over x^\mu(1+x^\nu)^\lambda } .
\end{equation}
%%%%%%%%%%%%%%%%%%%%%%%%%%%%%%%%%%%%%%%%%%%%%%%%%%%%%%%%%%%%%%%%%%

(ii) fix $B_{\rm p}$ (eq.[\ref{eq:bp}]), $R$ (eq.[\ref{eq:grdef}]) and
the polytropic exponent $n$.

(iii) solve the following equation for $\epsilon(x)$
%%%%%%%%%%%%%%%%%%%%%%%%%%%%%%%%%%%%%%%%%%%%%%%%%%%%%%%%%%%%%%%%%%
\begin{equation}
{d^2 \epsilon \over dx^2} + {2\over x} {d \epsilon \over dx}
+ B_{\rm p} \left[ {1 \over x^\mu(1+x^\nu)^\lambda } 
+ R\,\epsilon^n \right] = 0 ,
\end{equation}
%%%%%%%%%%%%%%%%%%%%%%%%%%%%%%%%%%%%%%%%%%%%%%%%%%%%%%%%%%%%%%%%%%
with the boundary condition at $x \ll 1$:
%%%%%%%%%%%%%%%%%%%%%%%%%%%%%%%%%%%%%%%%%%%%%%%%%%%%%%%%%%%%%%%%%%
\begin{equation}
\epsilon(x) = 1 - {B_{\rm p} \over (2-\mu)(3-\mu)}x^{2-\mu} .
\end{equation}
%%%%%%%%%%%%%%%%%%%%%%%%%%%%%%%%%%%%%%%%%%%%%%%%%%%%%%%%%%%%%%%%%%

(iv) then one obtains $T_{\rm g}(x) = T_{\rm g 0}\,\epsilon(x)$ and
$\rho_{\rm g}(x) = \rho_{\rm g 0}\,[\epsilon(x)]^n$, and finally can
compute the corresponding X-ray surface brightness $\Sigma_{\rm
X}(\theta)$ which should be compared with an appropriate sample of
X-ray clusters.

(v) repeat the above procedure to determine the set of parameters 
$\mu$, $\nu$, $\lambda$, $n$, $B_p$, and $R$. The latter two
quantities are combined to yield the amplitude of the halo density
$\delta_c\rho_{c0}$. 

With the current and future spatially resolved X-ray surface
brightness and temperature profiles of several clusters of galaxies
with Einstein, ROSAT, ASCA, AXAF, and XMM, the present methodology
will be a useful tool in revealing the shape of the underlying dark
matter halo potential which one has not been able to derive, and thus
provide important information on theories of cosmic structure
formation.

\bigskip

We thank the editor, E.L. Wright, for detailed and invaluable comments
on the earlier manuscript. This research was supported in part by the
Grants-in-Aid for the Center-of-Excellence (COE) Research of the
Ministry of Education, Science, Sports and Culture of Japan (07CE2002)
to RESCEU (Research Center for the Early Universe), University of
Tokyo, Japan.

\newpage
\bigskip
\bigskip

%%%%%%%%%%%%%%%%%%%%%%%%%%%%%%%%%%%%%%%%%%%%%%%%%%%%%%%%%%%%%%%%%%%%
%\baselineskip=12pt
%\parskip2pt
\centerline{\bf REFERENCES}
\bigskip

\def\apjpap#1;#2;#3;#4; {\pp#1, {#2}, {#3}, #4}
\def\apjbook#1;#2;#3;#4; {\pp#1, {#2} (#3: #4)}
\def\apjppt#1;#2; {\pp#1, #2.}
\def\apjproc#1;#2;#3;#4;#5;#6; {\pp#1, {#2} #3, (#4: #5), #6}

\apjpap Bardelli, S., Zucca, E., Malizia, A., Zamorani, G.,
Scaramella, R., \& Vettolani, G. 1996;A\&A;305;435;
\apjpap Bryan, G.L. \& Norman, M.L. 1998;ApJ;495;80;
\apjpap Evans, N.W. \& Collett, J.L. 1997;ApJ;480;L103;
\apjpap Fukushige, T., \& Makino, J. 1997;ApJ;477;L9;
\apjpap Ikebe et al. 1997;ApJ;481;660;
\apjppt Makino, N. \& Asano, K. 1998;submitted to ApJ;
\apjpap Makino, N., Sasaki, S., \& Suto, Y. 1998;ApJ;497;555 (Paper I);
\apjppt Markevitch, M., Forman, W.R., Sarazin, C.L., \& Vikhlinin, A.
1998;ApJ, in press (astro-ph/9711289);
\apjpap Moore, B., Governato, F., Quinn, T., Stadel, J., \& Lake, G.
1998;ApJ;499;L5;
\apjpap Navarro, J.F., Frenk, C.S., \& White, S.D.M. 1996;ApJ;462;
   563;
\apjpap Navarro, J.F., Frenk, C.S., \& White, S.D.M. 1997;ApJ;490;
  493;
\apjpap Melott,A.L,  Splinter,R.J., Shandarin, S.F., \&  Suto, Y. 1997; 
ApJ;479;L79;
\apjpap Splinter,R.J., Melott,A.L,   Shandarin, S.F., \&  Suto, Y. 1998; 
ApJ;497;38;
\apjppt Tamura, T. 1998;ph.D thesis (University of Tokyo), unpublished;
\apjpap White, S.D.M., Navarro, J.F., Evrard, A.E., \& Frenk, C.S.
1993;Nature;366;429;
\apjpap Xu, H., Makishima, K., Fukazawa, Y., Ikebe, Y., Kikuchi, K.,
Ohasi, T., \& Tamura, T. 1998;ApJ;500;738;
\apjpap Yoshikawa, K., Itoh, M. \& Suto, Y. 1998;PASJ;50;203;
%%%%%%%%%%%%%%%%%%%%%%%%%%%%%%%%%%%%%%%%%%%%%%%%%%%%%%%%%%%%%%%%%%%%%

%%%%%%%%%%%%%%%%%%%%%%%%%%%%%%%%%%%%%%%%%%%%%%%%%%%%%%%%%%%%%%%%%%%%%
%%%%%%%%%%%%%%%%%%%%%%%%%%%%%%%%%%%%%%%%%%%%%%%%%%%%%%%%%%%%%%%%%%%%%

%%%%%%%%%%%%%%%%%%%%%%%%%%%%%%%%%%%%%%%%%%%%%%%%%%%%%%%%%%%%%%%%%%%%%
%%%%%%%%%%%%%%%%%%%%%%%%%%%%%%%%%%%%%%%%%%%%%%%%%%%%%%%%%%%%%%%%%%%%%

%%%%%%%%%%%%%%%%%%%%%%%%%%%%%%%%%%%%%%%%%%%%%%%%%%%%%%%%%%%%%%%%%%%%%
%%%%%%%%%%%%%%%%%%%%%%%%%%%%%%%%%%%%%%%%%%%%%%%%%%%%%%%%%%%%%%%%%%%%%

\clearpage

%effect of gas gravity
%%%%%%%%%%%%%%%%%%%%%%%%%%%%%%%%%%%%%%%%%%%%%%%%%%%%%%%%%%%%%%%%%%%%%
%%%%%%%%%%%%%%%%%%%%%%%%%%%%%%%%%%%%%%%%%%%%%%%%%%%%%%%%%%%%%%%%%%%%%

%%%%%%%%%%%%%%%%%%%%%%%%%%%%%%%%%%%%%%%%%%%%%%%%%%%%%%%%%%%%%%%%%%%%%
%%%%%%%%%%%%%%%%%%%%%%%%%%%%%%%%%%%%%%%%%%%%%%%%%%%%%%%%%%%%%%%%%%%%%

%%%%%%%%%%%%%%%%%%%%%%%%%%%%%%%%%%%%%%%%%%%%%%%%%%%%%%%%%%%%%%%%%%%%%
%%%%%%%%%%%%%%%%%%%%%%%%%%%%%%%%%%%%%%%%%%%%%%%%%%%%%%%%%%%%%%%%%%%%%

%%%%%%%%%%%%%%%%%%%%%%%%%%%%%%%%%%%%%%%%%%%%%%%%%%%%%%%%%%%%%%%%%%%%%
%%%%%%%%%%%%%%%%%%%%%%%%%%%%%%%%%%%%%%%%%%%%%%%%%%%%%%%%%%%%%%%%%%%%%

\end{document}